\documentstyle[graphicx,multicol,prl,aps]{revtex}

\begin{document}

% \draft command makes PACS numbers print
 \draft

\title{Overlap Among States at Different Temperatures in the SK Model}

\author{Alain Billoire}

\address{
  Service de physique th\'eorique \\
  CEA Saclay,
  91191 Gif-sur-Yvette, France.
}

\author{Enzo Marinari}

\address{
  Dipartimento di Fisica, SMC and UdR1 of INFM and INFN, 
  Universit\`a di Roma {\em La Sapienza},\\
  P. A. Moro 2, 00185 Roma, Italy. }

\date{\today}                                                 

\maketitle

\begin{abstract}
We discuss the issue of temperature chaos in the
Sherrington--Kirkpatrick spin glass mean field model.  We numerically
compute probability distributions of the overlap among (equilibrium)
configurations at two different values of the temperature, both in the
spin glass phase.  The situation on our medium size systems is clearly
non-chaotic, but a weak form of chaos could be emerging on very large
lattices.
\end{abstract}


\begin{multicols}{2}
\narrowtext
\parskip=0cm

Two years after our paper \cite{FIRST} on the subject (that we will
quote as {\bf (A)} in the following) we are coming back to the problem
of {\em temperature chaos} in spin glasses and, more widely, in
disordered and complex systems. A problem that is, we believe, still a
very open one.

In these two years the problem of temperature chaos has been studied
under many new lights. One can fairly say that indications are mixed,
with some preference for a no-chaos scenario on medium size systems:
the detailed discussion of \cite{BOUCHA} (where arguments against the
possibility of a {\em strong chaos} picture are given) is probably a
perfect starting point for the reader interested in the 
details of the subject.

Perturbation theory in an expansion below $T_c$ (in mean field theory
\cite{RIZZO}) when pushed to the fifth order shows absence of chaos
through highly non-trivial cancelations, although one finds no
general feature that could imply that these cancelations will be
present at all orders in perturbation theory. The naive TAP equations
for the mean field Sherrington--Kirkpatrick model, when solved
numerically on systems with order of $10^2$ spins, also lead to
exclude the presence of temperature chaos \cite{MUPAPA}.

Bouchaud and Sales have shown that there is no chaos in the REM model
(but if one sits exactly at $T_c$) \cite{BOUSAL}. On the contrary
Sales and Yoshino have discussed in \cite{SALYOS} the case of DPRM
(directed polymers in random media), and have shown that there is
temperature chaos in this model, and that a temperature perturbation
plays a role very similar for example to a perturbation in the
potential: i.e.\ in the case of DPRM temperature perturbation looks
generic and creates chaos. The recent work by Sasaki and Martin 
\cite{SASMAR} describes situations where chaos is present.

These recent studies add new elements to the many former studies of an
interesting problem first discussed by Parisi \cite{PARISI}, and then
studied in many other works (see among
others\cite{KONDOR,NEYHIL,KONVEG,RITORT,FRANEY,NEYNIF}).  Still the
presence or the absence of temperature chaos is one of the few open
problems remaining in the Sherrington--Kirkpatrick model.

Let us start by reminding which were the main results of our first
paper {\bf (A)}. There we discussed the behavior of the
two-temperature overlap
\begin{equation}
  q_{T_1,T_2}^{(2),(N)} 
    \equiv 
  \overline{
    \left\langle 
      \left(
        \frac1{N}\sum_{i=1}^N \sigma_i^{T_1}\sigma_i^{T_2}
      \right)^2
    \right\rangle}\,
  \label{Q2}
\end{equation}
for systems with $N$ spins. The over-line is for the average over the
quenched disorder, the brackets are for the thermal average,
$\{\sigma_i^{T_1}\}$ is an equilibrium configuration of Ising spins at
temperature $T_1$, while $\{\sigma_i^{T_2}\}$ is at temperature $T_2$.
We have considered the Sherrington--Kirkpatrick spin glass mean field
model, a diluted finite connectivity mean field spin glass model, and
the $3D$ Edwards--Anderson spin glass. In a $T$-chaotic situation we
would expect $q_{T_1,T_2}^{(2),(N)}$ to go to zero in the infinite
volume limit, as soon as $T_1\ne T_2$. We found however that for all
models we studied (on systems with up to $4096$ sites)
$q_{T_1,T_2}^{(2),(N)}$ was not small for $T_2-T_1$ finite and
reasonably large.  To be more precise we found that in our data we had
always
\begin{equation}
  q_{T_1^{(min)},T_2}^{(2),(N)} 
  -
  q_{T_2,T_2}^{(2),(N)} > 0, \mbox{ for } 
  T_c \ge T_2 > T_1^{(min)} \ge 0.4 T_c\ ,
  \label{INEQ}
\end{equation}
even if the value of the l.h.s.\ was decreasing with increasing
volumes.  The validity of the relation (\ref{INEQ}) is suggestive of a
very non-chaotic situation (it would hold for an usual ferromagnet),
even if from the numerical data of {\bf (A)} it was clear that
asymptotically it may well be violated (that is in any case not enough
to imply a $T$-chaotic behavior).

So {\bf (A)} was suggesting a clear absence of chaotic behavior on
medium size systems, while showing that for increasing lattice sizes
signs of a (modestly) more chaotic behavior were possibly starting to
appear. We note here again that systems with a not huge number of
spins can be relevant to the physics of spin glasses. The experiments
of reference \cite{JOWHV} show that the number of spin involved in the
collective behavior observed in a typical spin glass experiment is of
the order of $10^5$, that is not so far away from the number of spins
we can handle numerically today. Accordingly, even if it should emerge
that $T$-chaos is asymptotically present for very large volumes, the
present numerical simulations could turn out to describe reasonably
well the experimental regime.

In order to get further informations on the sensitivity of the spin
glass phase with respect to temperature, we have decided to measure
the full probability distribution of the overlap of configurations
equilibrated at different values of the temperature: $P_{T_1,T_2}(q)$
(In {(\bf A)} we restricted ourselves to the second moment of this
distribution). In terms of  $P_{T_1,T_2}(q)$, temperature chaos just
means that $\lim_{N\to\infty}  P_{T_1,T_2}(q) = \delta(q)$, for any
$T_1\neq T_2$ .
A more complete information like the one contained in
the full probability distribution of the order parameter can indeed
allow a more detailed analysis of the scaling behavior, making
possible to uncover a wider range of phenomena.

Since the three models studied in {\bf (A)} were showing a very
similar behavior, we focus here on one of them, namely the
Sherrington--Kirkpatrick model.

The numerics are very similar to the ones of {\bf (A)}, and we refer
to \cite{FIRST} for a discussion of the details of the simulations.
We use binary quenched random couplings, $J=\pm 1$.  Here we have
studied systems with $N=64$, $N=256$, $N=1024$ and $N=4096$ spins,
down to $T=0.4=0.4\ T_c$.  We have used a multi spin (different spins
are encoded in the same computer word) version\cite{MSC} of the
parallel tempering Monte Carlo algorithm. Two copies of the system at
each one of a set of temperature values ($0.4, 0.4+\Delta T, 0.4+2
\Delta T \cdots$, with $\Delta T=0.025$ but for $N=4096$ where $\Delta
T=0.0125$) are brought to equilibrium (in the same realization $J$ of
the couplings) and are used to compute the $P_{T_i,T_j}^{(J)}(q)$
(more precisely the subset with temperatures $0.40, 0.45, 0.50,
\cdots$) .  An average over $1024$ different realizations of the
quenched disorder ($256$ for $N=4096$) is taken to compute the average
$P_{T_i,T_j}(q)$. This is an order of magnitude more disorder
samples than in  \cite{FIRST}. For each
realization we have performed $400 K$ sweeps for equilibrium plus
$1000 K$ sweeps for measurements ($520 K$ only for $N=4096$).

\begin{figure}
  \centering
  \includegraphics[width=0.33\textwidth,angle=270]{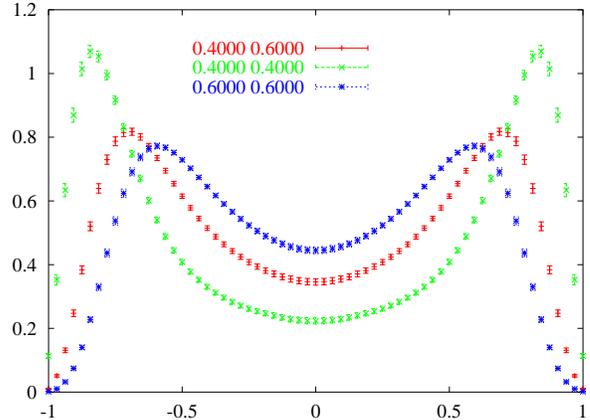}
  \caption[a]{
    $P(q)$ in a non-diagonal case 
    ($T_1=0.4$ and $T_2=0.6$) 
    and for the two corresponding diagonal cases
    ($T_1=T_2=0.4$
    and $T_1=T_2=0.6$). 
    Here $N=64$.}
  \protect\label{F-PQ64}
\end{figure}

\begin{figure}
  \centering
  \includegraphics[width=0.33\textwidth,angle=270]{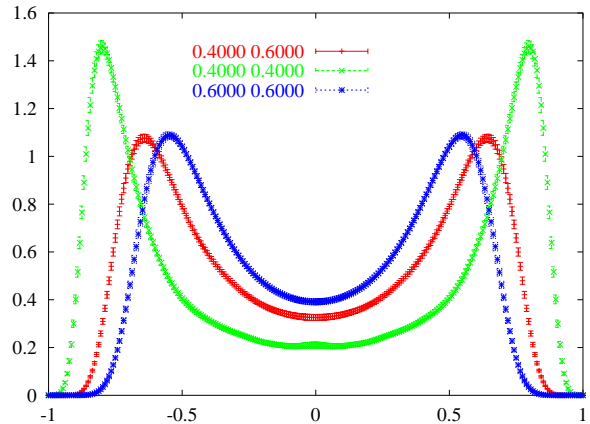}
  \caption[a]{
    As in figure \ref{F-PQ64} but for $N=256$.
  }
  \protect\label{F-PQ256}
\end{figure}

\begin{figure}
  \centering
  \includegraphics[width=0.33\textwidth,angle=270]{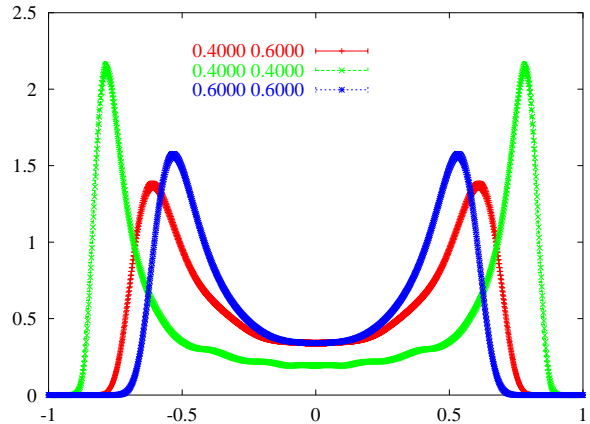}
  \caption[a]{
    As in figure \ref{F-PQ64} but for $N=1024$.
  }
  \protect\label{F-PQ1024}
\end{figure}

\begin{figure}
  \centering
  \includegraphics[width=0.33\textwidth,angle=270]{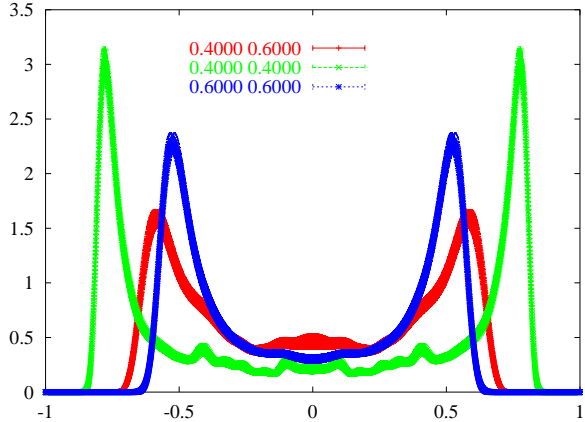}
  \caption[a]{
    As in figure \ref{F-PQ64} but for $N=4096$.
  }
  \protect\label{F-PQ4096}
\end{figure}

We show at first the probability distributions themselves (normalized
by $\int^1_{-1} P(q) dq =1$). This distributions have been symmetrized,
although they are quite symmetric, even sample by sample.
In each of figures \ref{F-PQ64},
\ref{F-PQ256}, \ref{F-PQ1024} and \ref{F-PQ4096} we show three
$P_{T_1,T_2}(q)$: the non-diagonal one that gives the probability of
the overlap of configurations at $T=0.4$ (the lower temperature we
equilibrate) with configurations at $T=0.6$, and the two diagonal
probability distributions with $T_1=T_2=0.4$ and $T_1=T_2=0.6$
respectively.  In figure \ref{F-PQ64} we have data from our smallest
lattice with $N=64$, in figure \ref{F-PQ256} from $N=256$, in figure
\ref{F-PQ1024} from $N=1024$, and in figure \ref{F-PQ4096} we have
data from the largest lattice, with $N=4096$.

On the smallest lattice (figure \ref{F-PQ64}) the peak of the
non-diagonal $P(q)$ is higher than the one of $P_{0.6,0.6}$, and its
position is basically at half way between the positions of the maxima
of $P_{0.4,0.4}$ and $P_{0.6,0.6}$. This is typically what would
happen in a ferromagnet. For increasing volumes the position of this
peak is moving (very slowly) to lower values of $q$, but does not seem
to approach $q=0$ (we will discuss in detail this point in
the following).

\begin{figure}
  \centering
  \includegraphics[width=0.33\textwidth,angle=270]{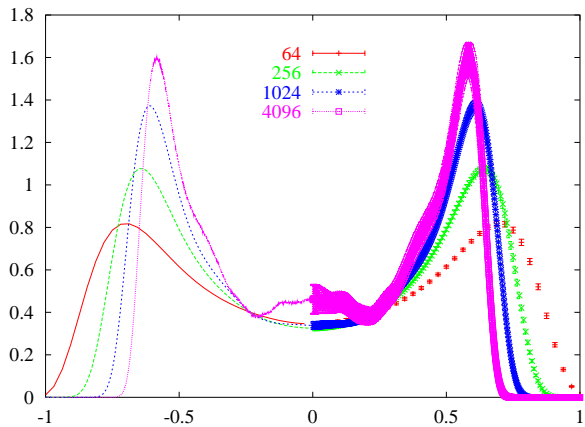}
  \caption[a]{$P(q)$  for the non-diagonal case only
    ($T_1=0.4$ and $T_2=0.6$) for different lattice volumes.
For giving more information about the actual form of the measure curves
we only plot errors in the $q>0$ part of the plot.}
  \protect\label{F-PQNODIAG}
\end{figure}

\begin{figure}
  \centering
  \includegraphics[width=0.33\textwidth,angle=270]{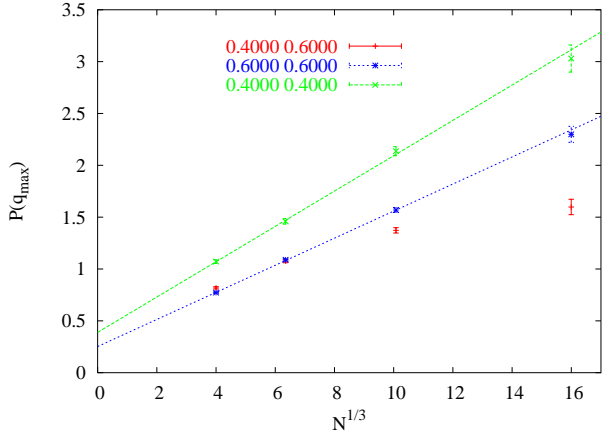}
  \caption[a]{$P(q_{\mbox{max}})$ 
as a function of $N^{\frac13}$
for   the three different $P(q)$.}
  \protect\label{F-MAXBB}
\end{figure}

In figure \ref{F-PQNODIAG} we show in the same plot the four
non-diagonal $P_{0.4,0.6}$ for $N=64$, $256$, $1024$ and $4096$.
These $P_{0.4,0.6}(q)$ are normalized and are drawn on the same scale,
so that a visual comparison of the four curves is meaningful.  Here we
see that the $q\ne 0$ peak of $P_{0.4,0.6}(q)$ do increase as function
of $N$ (but for a negative analysis of this statement see later in the
text and figure \ref{F-MAXBB}), and that its position shifts only very
little towards $q=0$. The mass carried by the distribution $P(q)$ in
$q\approx 0$ seems to be increasing  on the largest lattice,
but this effect is affected by a large statistical error.  Figure
\ref{F-PQ4096} shows indeed wiggles in $P_{0.4,0.4}(q)$ that are known
not to exist in the infinite volume limit, and are accordingly to be
attributed to the limited number of disorder samples.
The bump popping up around $q=0$ in figure \ref{F-PQNODIAG} 
is  compatible with the statistical error. It is not visible in our data
for larger $T_1$ values ($0.45, 0.50,\cdots$).

\begin{figure}
  \centering
  \includegraphics[width=0.33\textwidth,angle=270]{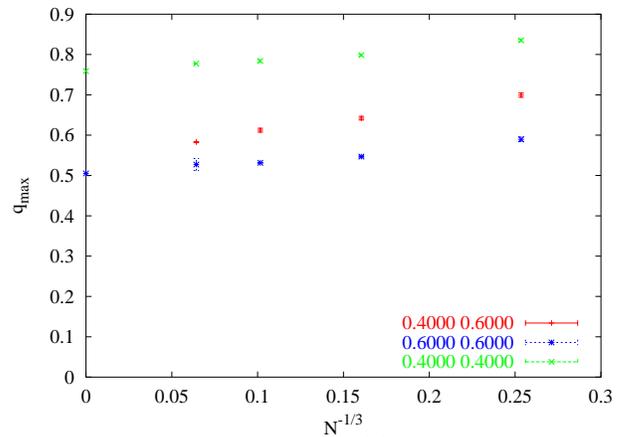}
  \caption[a]{The values of $q$ where $P(q)$ is
maximum, $q_{\mbox{max}}$, for the diagonal and non-diagonal $P(q)$ 
as function of $N^{-1/3}$.}
  \protect\label{F-LOC}
\end{figure}

In figure \ref{F-LOC} we plot the values of $q$ where $P(q)$ is
maximum, $q_{\mbox{max}}$, for the diagonal and non-diagonal $P(q)$ at
different $N$ values. The two points at $N=\infty$ for
$P_{0.4,0.4}(q)$ and $P_{0.6,0.6}(q)$ have been obtained \cite{CRIRIZ}
using the method of \cite{CRI-RIZ}: they show how reliable is a linear
fit in $N^{-\frac13}$ of our data for the diagonal distributions. The
points for $P_{0.4,0.6}(q)$ show some curvature, but the effect is not
dramatic, and a non-zero value in the limit $N\to\infty$ is surely
favored.

As we have already shown, the peaks of the (temperature) non-diagonal
$P(q)$ increase when $N$ increases, but not as fast as the ones of
the diagonal $P(q)$. In figure \ref{F-MAXBB} we show the value of
$P(q_{\mbox{max}})$ for the different $N$ values and for the three
$P(q)$ we are discussing. The height of the peaks of the two diagonal
$P(q)$ increases exactly like $N^{\frac13}$ as it should.  The two
continuous straight lines are  the best linear fit to the diagonal
data: the fits turn out to be very good.

From figure \ref{F-MAXBB} the situation of $P_{0.4,0.6}(q)$ looks
quite different. Here the growth slows down at $N=1024$ and is really
small on our largest lattice size with $N=4096$.  We do not believe
that one can draw precise quantitative conclusions from figure
\ref{F-MAXBB}: a (weak) non-chaotic picture could survive in the
infinite volume limit, or chaos could appear very slowly (only on very
large volumes).  Let us spell clearly, in any case, that the mechanism
that will govern a possible appearance of chaos will be based on the
$P_{T_1,T_2}(q)$ having two $q=0$ and and at $q=q^*>0$, and by having
the $q=0$ peak growing for increasing volume at the expense of the
$q^*$ peak.  In any case our numerical results appear to support the
idea that temperature chaos will not be effective on the typical sizes
that are relevant in spin glass experiments \cite{JOWHV}.

\begin{figure}
  \centering
  \includegraphics[width=0.33\textwidth,angle=270]{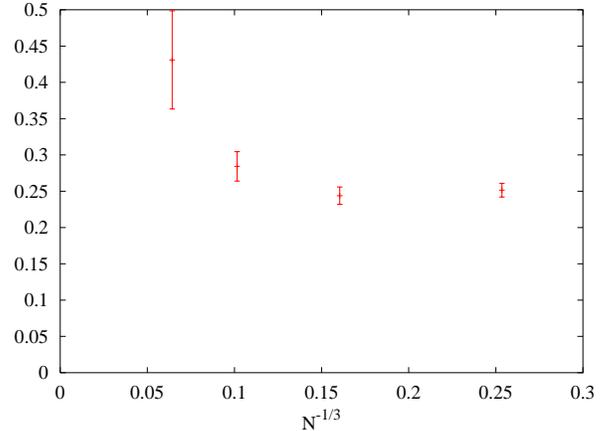}
  \caption[a]{${\cal R}$ as defined in equation (\ref{ERRE}) as a
  function of  $N^{-\frac13}$.}
  \protect\label{F-RATIO}
\end{figure}

In order to  be more quantitative we have also looked at the
ratio of the mass carried by  $P_{0.4,0.6}(q)$ close to $q=0$ as
compared to the mass carried by the large $q$ region. We define

\begin{equation}
  {\cal R} \equiv k\ \frac{\int_{-m}^{+m}
  P(q)\ dq}{\int^1_{q_{{max}}}P(q)\ dq} \ ,
  \label{ERRE}
\end{equation}
where $k$ is a normalization constant, and we take $m=0.05$ (a very
similar picture would be obtained for any $m$ not too large).  $\cal
R$ decreases with $N$ if the mass of the peaks at large $q$ dominates,
while it increases when the dominating contribution is the one at
$q\approx 0$.

We plot $\cal R$ for $P_{0.4,0.6}(q)$ versus $N^{-\frac13}$ in figure
\ref{F-RATIO}. $\cal R$ is constant on the smaller lattices, but
starts increasing on the largest lattice size. The error due to sample
to sample fluctuations is here very large (the simulation at $N=4096$
have been very costly in computer time), and the growth of $\cal R$ is
not significant at more than two standard deviations, but an effect is
very plausibly there. Again, this is probably an indication toward the
fact that temperature chaos could eventually emerge on very large
systems. 

It is worth noticing that if temperature chaos is (slowly) emerging
when increasing $N$ this is probably not happening with the position
of the two peaks at large $|q|$ shifting to $q=0$, but with a third
peak in $q=0$ emerging and eventually becoming the only contribution,
and accordingly a discontinuity in $q_{\mbox{max}}$ as function of
$N^{-1/3}$ (Figure \ref{F-LOC}).

The problem of temperature chaos, already at the mean field level, is
turning out to be a hard problem: this is true both for analytical and
for numerical computations. Here we have provided some further hints
about the behavior of the system in the infinite volume limit: the very
large scale, state of the art simulations we have been able to run,
give some suggestions, probably hinting in favor of a very weak
chaos that would emerge only for very large lattice sizes. 

We acknowledge many enlightening discussions with Giorgio Parisi and
Silvio Franz.  We thank Andrea Crisanti and Tommaso Rizzo for
providing us with the $N=\infty$ data for $q_{\mbox{max}}$ for the
diagonal $P(q)$ of figure \ref{F-LOC}. The numerical simulations have
been run on the CEA Compaq Alpha Server at Grenoble.

\end{multicols}
\end{document}